# Switching of magnetization by non-linear resonance studied in single nanoparticles

Christophe Thirion[1], Wolfgang Wernsdorfer[1], Dominique Mailly[2]

[1] Laboratoire Louis Néel, associé à l'UJF, CNRS, BP 166, 38042 Grenoble Cedex 9, France
[2] Laboratoire de photonique et de nanostructures, CNRS, Route de Nozay, 91460 Marcoussis, France

Magnetization reversal in magnetic particles is one of the fundamental issues in magnetic data storage. Technological improvements require the understanding of dynamical magnetization reversal processes at nanosecond time scales.[1] New strategies are needed to overcome current limitations. For example, the problem of thermal stability of the magnetization state (superparamagnetic limit) can be pushed down to smaller particle sizes by increasing the magnetic anisotropy.[2] High fields are then needed to reverse the magnetization that are difficult to achieve in current devices. Here we propose a new method to overcome this limitation. A constant applied field, well below the switching field, combined with a radio-frequency (RF) field pulse can reverse the magnetization of a nanoparticle. The efficiency of this method is demonstrated on a 20 nm cobalt particle by using the micro-SQUID technique.[3] Other applications of this method might be nucleation or depinning of domain walls.

The dynamics of magnetization reversal of magnetic particles is of great relevance for applications in spintronic circuits operating in the GHz range. Recent improvements of current limitations concern the magnetization switching via precessional modes overcoming the relaxation time limit.[4-9] It has also been shown that a spin polarized current can be used to switch the magnetization state in nanometer sized dots.[10] The method presented here aims to reduce magnetization switching fields of nanoparticles. Apart from technological applications, our technique is of particular importance in view of theoretical predictions[11-13] because the experimental determination of the most effective frequency is a direct probe of the precession of the magnetization in the metastable energy well. This allows to access to the precessional damping being a key issue for all magnetization processes at nanosecond time scales.

The magnetization reversal of a single-domain nanoparticle can be described by the Stoner-Wohlfarth model.[14] One considers a particle of an ideal magnetic material where exchange energy holds all spins tightly parallel to each other. There is competition only between the anisotropy energy of the particle and the effect of the applied field. The reversal of the magnetization is described by the potential energy:[15]

$$E(\vec{M}, \vec{H}) = E_0(\vec{M}) - \mu_0 V \vec{M} \vec{H} \qquad (1)$$

where $V$ and $\vec{M}$ are the magnetic volume and the magnetization vector of the particle respectively, $\vec{H}$ is the external magnetic field, and $E_0(\vec{M})$ can be an arbitrary effective anisotropy which includes shape, magnetocrystalline, magnetoelastic and surface anisotropies. In the case of uniaxial anisotropy, the energy potential has two wells corresponding to the two stable orientations of the magnetization (Fig. 1a). When a field is applied, one of the two wells becomes metastable. At a particular field, called the switching field, the energy barrier between the two wells vanishes and the magnetization reverses. Fig. 1b shows the switching field as a function of the direction of the applied field. The corresponding curve is called the Stoner-Wohlfarth astroid.[14] This model was very successful because of its simplicity. However, it does not describe the reversal path. The time dependence of the magnetization $\vec{M}$ under the influence of an effective magnetic field $\vec{H}_{eff}$, is described be the Landau-Lifschitz-Gilbert (LLG) equation:[1]

$$\frac{d\vec{M}}{dt} = -\gamma \vec{M} \times \vec{H}_{eff} + \frac{\alpha}{|\vec{M}|} \vec{M} \times \frac{d\vec{M}}{dt} \qquad (2)$$

The first term on the right-hand side is the precession term and the second one is a phenomenological damping term. The effective field $\vec{H}_{eff}$ is defined as the sum of all fields acting on the magnetization, that is the applied field $\vec{H}$ and the anisotropy field. $\vec{H}_{eff}$ is the derivative of Eq. 1 with respect to $\vec{M}$, that is $\vec{H}_{eff} = -\frac{1}{\mu_0 V} \frac{\partial E}{\partial \vec{M}}$. The damping term allows the magnetization to reach the minimum of a potential well. Our method of magnetization reversal using non-linear resonance consists in driving the precession of magnetization with an applied RF field pulse until the magnetization switches (Fig. 1a). The method proved to be very efficient when the precession frequency matches the frequency of the RF pulse. A tiny RF pulse amplitude is sufficient to reduce strongly the switching fields.

We studied this reversal method on individual nanoparticles by using planar Nb micro-bridge-DC-SQUIDs.[3,16] The cobalt nanoparticles encapsulated in graphitic cages were synthesized using an arc-discharge method.[17] They are monocrystals of pure cobalt and have a face-centered cubic (fcc) or a hexagonal close-packed (hcp) structure.[18] The presented data were obtained on a hcp particle with a diameter of about 20 nm.

The switching fields of the magnetization of single Co nanoparticles were measured by scanning the applied field in all directions of space at a rate of about 0.1 T/s (fig. 3). Experimental details are described elsewhere.[3] This switching field map characterizes completely the magnetic anisotropy function $E_0(\vec{M})$ in Eq. (1)  [19] and is therefore very important for simulating the dynamics (Eq. (2)).

We use the micro-SQUID like a strip line to apply the RF pulse to the nanoparticle (see method section and Fig. 2b). Typical results are presented in Fig. 4a for several RF frequencies. For all fields, which are inside a given curve, the magnetization does not switch during the RF pulse. The black curve shows the switching field without field pulse. For all fields between these curves, the magnetization reversal is triggered by the RF pulse. In particular field regions, the switching field is strongly reduced by the RF pulse. Figs. 4b and c present an enlargement of the most sensitive field regions as a function of pulse length. A RF pulse of about 1 ns can already be sufficient to reverse the magnetization. The reason for this selective sensitivity can be understood by recalling that the RF pulse frequency has to match the precession frequency of the particle that depends strongly on the applied field.

A qualitative understanding of the magnetization reversal by non-linear resonance can be obtained with the LLG equation (Eq. 2), which can be solved numerically using a modified Runge-Kutta integration scheme. The initial state of the magnetization is in the metastable minimum. Then, a RF pulse with a smooth envelope is applied, inducing a precession of the magnetization in the minimum. When the RF frequency matches the precession frequency, the magnetization spirals up to the saddle point and reverses (fig. 1a). We define the switching time as the time interval between the start of the RF pulse and the time when the $M_z$ components of the magnetization crosses zero. Fig. 5 presents the calculated switching times for all fields lying inside the Stoner-Wohlfarth astroid. In the black region, the RF pulse does not switch the magnetization. In most of all other cases, the switching time is around 1 ns but in few cases it can reach up to 3 ns.

A detailed numerical study of the magnetization reversal via non-linear resonance shows all features measured on the Co nanoparticle (fig. 4). The main features are (i) the resonance region shifts towards the hard axis for increasing RF frequency; (ii) the optimum damping constant $\alpha$ lies between 0.001 and 0.05; (iii) a small RF amplitude is sufficient to reduce strongly the switching field. In the case of Fig. 4c, we achieved a switching field reduction of about 100 mT with a RF amplitude of few mT (at 4.4 GHz). Even higher switching field reduction could be achieved by changing the pulse frequency during the RF pulse. Apart of technological applications, this technique probes directly the dissipative damping of the precession in the metastable well. The width of the resonance and the power required to achieve reversal are related to the dissipative damping of the oscillations and could therefore be used to determine the damping constant on a single nanoparticle.

Future measurements will focus on the effects induced by surface anisotropy and temperature on the magnetization switching via non-linear resonance. Other applications of our method might be the nucleation of magnetization reversal in magnetic dots or the depinning of domain walls in magnetic nanostructures. Even for magnetization reversal with spin-polarized current[10], a RF field or current pulse can favor reversal and probe the presessional modes. We expect therefore that our method become an important tool to probe magnetization reversal dynamics.

**Method section.**

In order to place one nanoparticle on the SQUID detector, we disperse the particles in ethanol by ultrasonication. Then we place a drop of this liquid on a chip of about hundred SQUIDs. When the drop is dry the nanoparticles stick on the chip due to Van der Waals forces. Only in the case when a nanoparticle falls on a micro-bridge of the SQUID loop, the flux coupling between SQUID loop and nanoparticle is strong enough for our measurements. Finally we determine the exact position and shape of the nanoparticles by scanning electron microscopy. In some cases, we used an atomic force microscope to reposition the nanoparticles[20].

The RF pulses were generated with a frequency synthesizer (Anritsu MG3694A) triggered with a nanosecond pulse generator. They were injected into the superconducting leads of the micro-SQUID via appropriate induction loops. The generated RF supercurrent goes through the micro-bridge junction of the SQUID and induces a RF field that is directly coupled to the nanoparticle (Fig. 2). The RF field direction can be found be measuring the 3D switching field astroid (Fig. 3) with and without a static current through the micro-bridge. The static current shifts the entire field astroid. The shift direction gives the RF field direction and the shift amplitude gives the field-coupling factor. Then, we used the micro-SQUID to estimate the amplitude in the high-frequency regime. Knowing the field-coupling factor allows us to estimate the amplitude. Finally, the amplitude variations for different pulse lengths was corrected using a 50 GHz digital sampling oscilloscope.

Because we use the micro-SQUID like a strip line to apply the RF pulse to the particle, the micro-SQUID does not function properly during the RF pulse. We developed therefore an indirect method[19,21] that consists of four steps. First, the magnetization of the particle is saturated in a given direction. Second, a test field is applied that is lower than the static switching field. Third, a RF pulse of given frequency, amplitude, and length is applied to the particle. Finally, the SQUID is switched on in order to determine whether the particle has switched during the RF pulse. Then, the entire procedure is repeated with another test field, frequency, amplitude, or pulse length.

**References.**

1. Hillebrands, B. & Ounadjela, K. (eds.) *Spin dynamic in confined magnetic structures*. (Springer, Berlin, 2002).

2. Sun, S., Murray, C.B., Weller, D., Folks, L. & Moser A. Monodisperse FePt nanoparticles and ferromagnetic FePt nanocrystal superlattices. *Science* **287**, 1989-1992 (2000).

3. Wernsdorfer, W. Classical and quantum magnetization reversal studied in nanometer-sized particles and clusters. *Adv. Chem. Phys.* **118**, 99 (2001).

4. Back, C.H. et al. Minimum Field Strength in Precessional Magnetization Reversal. *Science* **285**, 864-867 (1999).

5. Gerrits, Th., van den Berg, H.A.M., Hohlfeld, J., Bär, L. & Rasing, Th. Ultrafast precessional magnetization reversal by picosecond magnetic field pulse shaping. *Nature* **418**, 509-511 (2002).

6. Kaka S. & Russek, S.E. Precessional switching of submicrometer spin-valves. *Appl. Phys. Lett.* **80**, 2958-2960 (2002).

7. Schumacher, H.W. et al. Phase coherent precessional magnetization reversal in microscopic spin valve elements. *Phys. Rev. Lett.* **90**, 17201-1-4 (2003).

8. Schumacher, H.W., Chappert, C., Sousa, R.C., Freitas, P.P. & Miltat, J. Quasiballistic magnetization reversal. *Phys. Rev. Lett.* **90**, 017204-1-4 (2003).

9. Hillebrands, B. & Fassbender, J. Ultrafast magnetic switching. *Nature* **418**, 493-494 (2002).

10. Myers, E.B., Ralph, D.C., Katine, J.A., Louie, R.N. & Buhrman, R.A. Current-Induced Switching of Domains in Magnetic Multilayer Devices. *Science* **285**, 867-870 (1999).

11. Garcia-Palacios, J.L. & Lazaro, F.J. Langevin-dynamics study of the dynamical properties of small magnetic particles. *Phys. Rev. B* **58**, 14937-14958 (1998).

12. Coffey, W.T. et al. Thermally activated relaxation time of a single domain ferromagnetic particle subjected to a uniform field at an oblique angle to the easy axis: Comparison with experimental observations. *Phys. Rev. Lett.* **80**, 5655-5658 (1998).

13. Bauer, M., Fassbender, J., Hillebrands, B. & Stamps, R.L. Switching behavior of a Stoner particle beyond the relaxation time limit *Phys. Rev. B* **61**, 3410-3416 (2000).

14. Stoner, E.C. & Wohlfarth, E.P. A mechanism of magnetic hysteresis in heterogeneous allows. *Philos. Trans. London Ser. A* **240**, 599-608 (1948).

15.     Thiaville, A. Coherent rotation of magnetization in three dimensions: A geometrical approach. *Phys. Rev. B* **61**, 12221-12232 (2000).

16.     Wernsdorfer, W. et al. Experimental evidence of the Néel-Brown model of magnetization reversal. *Phys. Rev. Lett.* **78**, 1791-1794 (1997).

17.     Guerret-Piécourt, C., Le Bouar, Y., Loiseau, A. & Pascard, H. Relation between metal electronic-structure and morphology of metal-compounds inside carbon nanotubes. *Nature* **372**, 761-765 (1994).

18.     Wernsdorfer, W., Thirion, C., Demoncy, N., Pascard, H. & Mailly, D. Magnetisation reversal by uniform rotation (Stoner-Wohlfarth model) in fcc cobalt nanoparticles. *J. Magn. Magn. Mat.* **242-245**, 132-137 (2002).

19.     Bonet, E. et al. Three-dimensional magnetization reversal measurements in nanoparticles. *Phys. Rev. Lett.* **83**, 4188-4191 (1999).

20.     Schleicher, B. et al. Magnetization reversal measurements of size-selected iron oxide particles produced via an aerosol route. *Applied Organometallic Chemistry* **12**, 315-320 (1998).

21.     Thirion, C. et al. Micro-SQUID technique for studying the temperature dependence of switching fields of single nanoparticles. *J. Magn. Magn. Mat.* **242–245**, 993-994 (2002).

**Acknowledgements**
This work was supported by the European Union network MASSDOTS. B. Barbara, A. Benoit, E. Bonet, and H. Pascard are acknowledged for the continous support of this research.

**Competing interests statement**
The authors declare that they have no competing financial interests

Correspondence and requests for materials should be addressed to W.W. (wernsdor@grenoble.cnrs.fr).

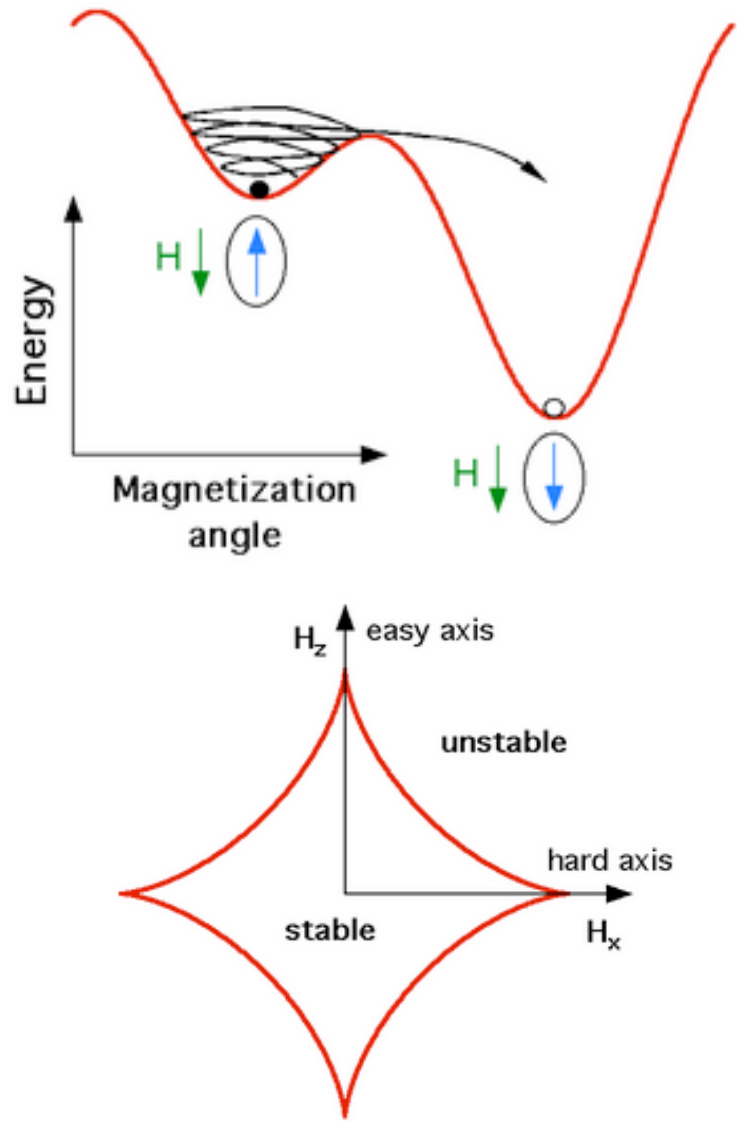

**Figure 1.**

Schematic representation of the magnetization reversal via non-linear resonance.

(a) Potential energy of a magnetic nanoparticle versus magnetization angle. The potential energy has two wells corresponding to two stable orientation of the magnetization. For small applied fields, one of the two wells is metastable. An additional applied alternative field $\delta H_{RF}$ induces oscillations of the magnetization in the energy wells. When the frequency of $\delta H_{RF}$ matches the precession frequency of the magnetization, energy can be pumped into the system. This can lead to magnetization reversal form the metastable to the stable well when the precessional damping is not too high.

(b) Field dependence of the switching field, called Stoner-Wohlfarth astroid. The easy and hard axes of magnetization are along the z- and x-directions, respectively. For all fields inside the astroid, the magnetization has two stable orientations (Fig. 1a) whereas there is only one outside. Therefore, the magnetization switches from the metastable well to the stable one when the applied field crosses the Stoner-Wohlfarth astroid.

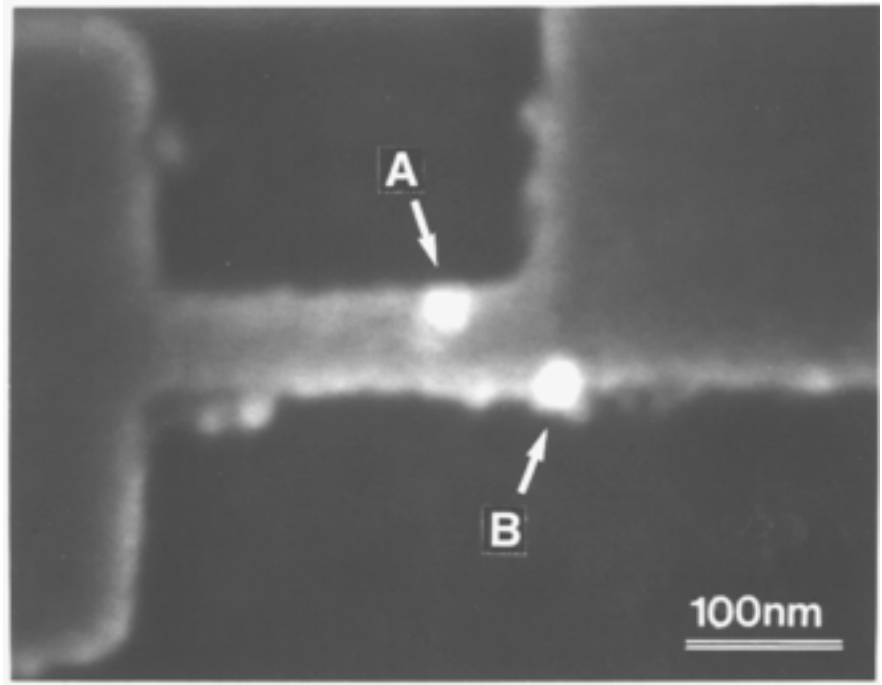

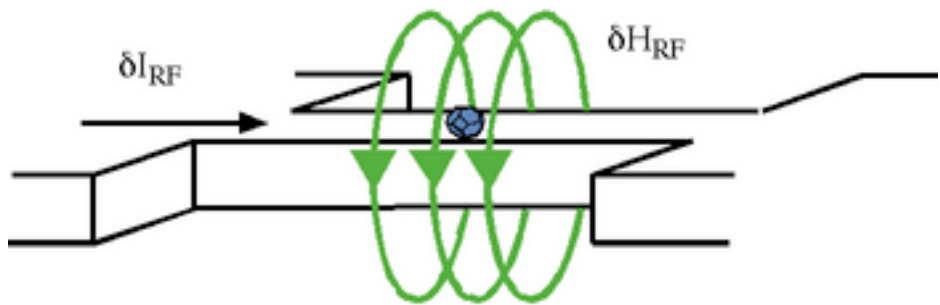

**Figure 2.**
Josephson junction (micro-bridge) of the micro-Superconducting QUantum Interference Device (SQUID) on which a 20 nm cobalt particle was placed.
(a) Scanning electron microscope image of the micro-bridge junction. The data presented here were obtained on particle A. The SQUID is patterned out of a 20 nm thick niobium film.
(b) Schematic view of the micro-bridge junction. The micro-bridge of the SQUID is used like a strip line. An injected RF supercurrent $\delta I_{RF}$ induces a RF field $\delta H_{RF}$ that is directly coupled to the nanoparticle on the micro-bridge.

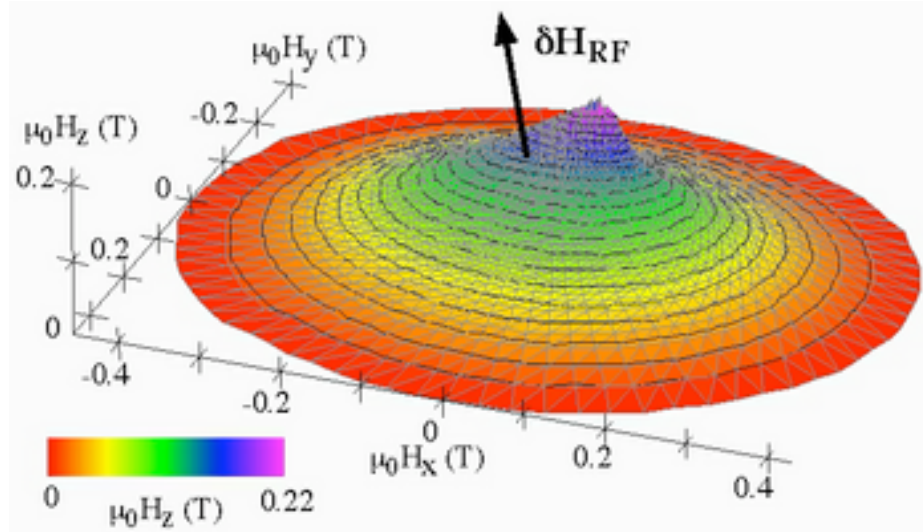

**Figure 3.**
Three-dimensional switching field map of the 20 nm Co particle shown in fig. 2a (particle A). The data can be described with the Stoner-Wohlfarth model using the potential energy (Eq. 1): $E_0\left(\vec{M}\right) = K_1 V \sin^2\theta + K_2 V \sin^4\theta$ where $K_1 V$ and $K_2 V$ are anisotropy energy constants taking into account of shape and magnetocrystalline anisotropy, and $\theta$ is the angle of magnetization with respect to the easy axis of magnetization. Note that the easy axis of magnetization is slightly tilted away from the z-direction and is not exactly in the xz-plane that is chosen in Fig. 4. The direction of the radio-frequency field pulse $\delta H_{\text{RF}}$ is indicated.

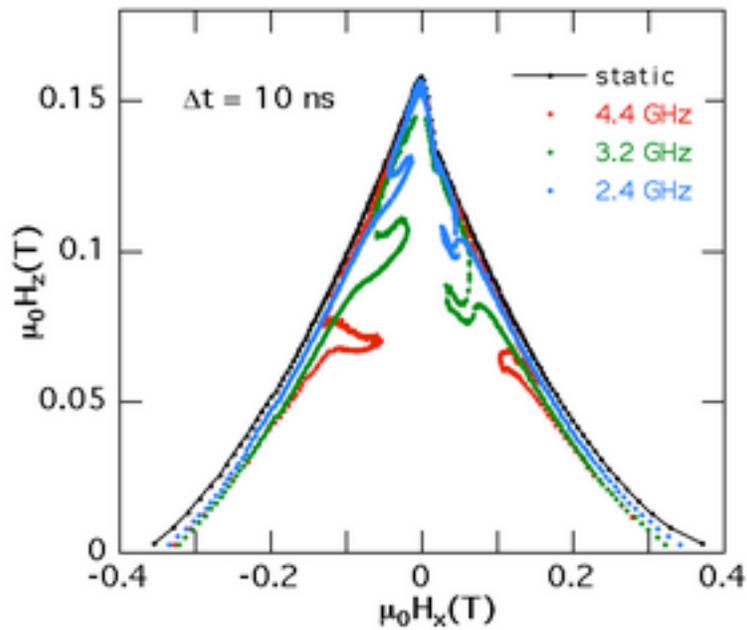

**Figure 4a**

**Figure 4.**
Field dependence of the switching field of the 20 nm Co particle (particle A in fig. 2a). The magnetic fields are applied in the xz-plane (Fig. 3).
(a) The black and color curves are the static and dynamic Stoner-Wohlfarth astroids, respectively. For the latter, the RF pulse frequencies are indicated and the pulse length is about 10 ns. Note that the effect is not symmetric because the RF field direction was not aligned with the easy (hard) axis of magnetization (Fig. 3).

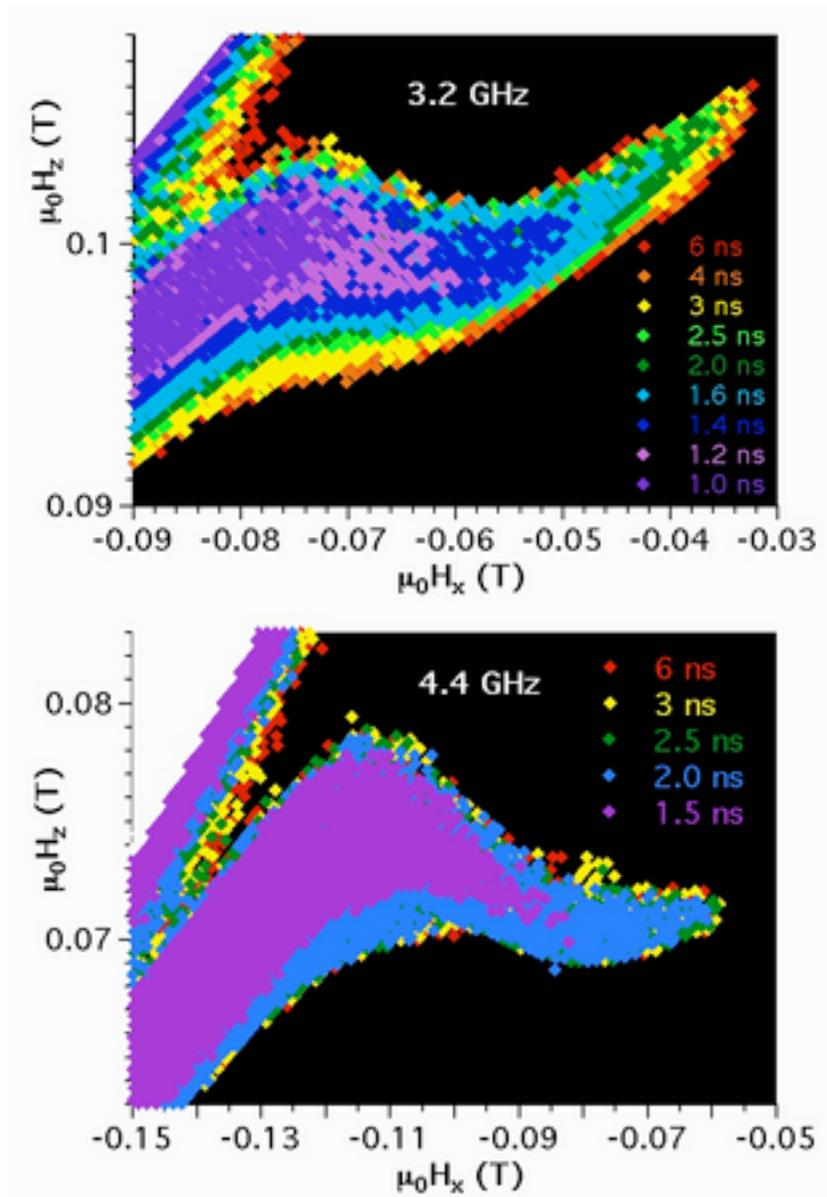

**Figure 4.**

Field dependence of the switching field of the 20 nm Co particle (particle A in fig. 2a). The magnetic fields are applied in the xz-plane (Fig. 3).

(a) The black and color curves are the static and dynamic Stoner-Wohlfarth astroids, respectively. For the latter, the RF pulse frequencies are indicated and the pulse length is about 10 ns. Note that the effect is not symmetric because the RF field direction was not aligned with the easy (hard) axis of magnetization (Fig. 3).

(b-c) Enlargement of the most sensitive field regions in Fig. 4a as a function of pulse length. For each field point, the shortest pulse length leading to magnetization switching is indicated with a color. In the black region the magnetization did not reverse whereas the white region is outside the Stoner-Wohlfarth astroid.

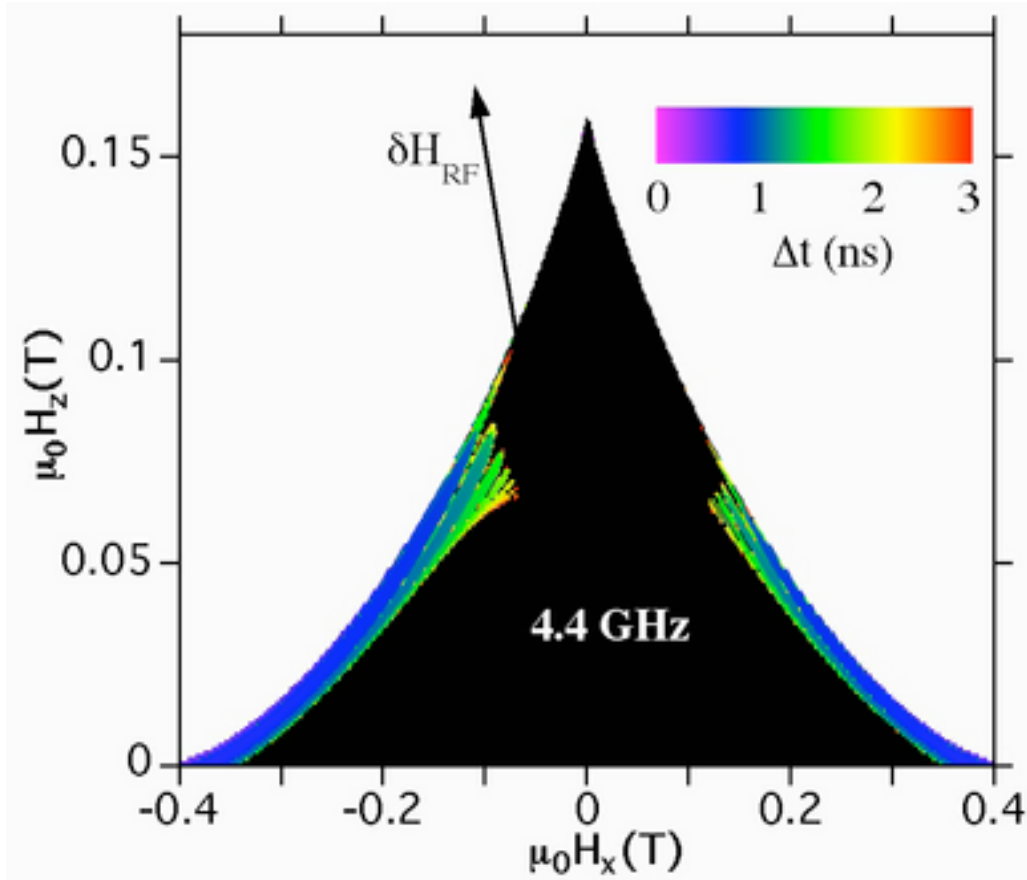

**Figure 5.**
Simulation of the field dependence of the switching field (dynamic Stoner-Wohlfarth astroid). For each field point inside the Stoner-Wohlfarth astroid, the LLG equation was integrated to calculate the switching times, which are plotted using the indicated color scale. The following parameters have been chosen: $M_S = 1.4 \times 10^6$ A/m (pure cobalt), $K_1/M_S = 0.23$ T, $K_2/M_S = 0.09$ T, $\alpha = 0.02$, and $\delta H_{RF} = 0.009$ T applied at an angle of 20° in respect to the easy axis of magnetization.